\documentclass[conference]{IEEEtran}

\usepackage{graphicx}

\begin{document}

\title{Exploiting Interference Alignment in Multi-Cell Cooperative
OFDMA Resource Allocation}

\author
{ \IEEEauthorblockN{Bin Da${}^\dagger$} \IEEEauthorblockA
{${}^\dagger$Department of ECE, National University of Singapore\\
Email:\{dabin,elezhang\}@nus.edu.sg} \and \IEEEauthorblockN{Rui
Zhang${}^\dagger{}^\ddagger$} \IEEEauthorblockA
{${}^\ddagger$Institute for Infocomm Research, A*STAR, Singapore\\
Email: rzhang@i2r.a-star.edu.sg} }

\maketitle

\begin{abstract}
This paper studies interference alignment (IA) based multi-cell
cooperative resource allocation for the downlink OFDMA with
universal frequency reuse. Unlike the traditional scheme that treats
subcarriers as separate dimensions for resource allocation, the IA
technique is utilized to enable frequency-domain precoding over
parallel subcarriers. In this paper, the joint optimization of
frequency-domain precoding via IA, subcarrier user selection and
power allocation is investigated for a cooperative three-cell OFDMA
system to maximize the downlink throughput. Numerical results for a
simplified symmetric channel setup reveal that the IA-based scheme
achieves notable throughput gains over the traditional scheme only
when the inter-cell interference link has a comparable strength as
the direct link, and the receiver SNR is sufficiently large.
Motivated by this observation, a practical hybrid scheme is proposed
for cellular systems with heterogenous channel conditions, where the
total spectrum is divided into two subbands, over which the IA-based
scheme and the traditional scheme are applied for resource
allocation to users located in the cell-intersection region and
cell-non-intersection region, respectively. It is shown that this
hybrid resource allocation scheme flexibly exploits the downlink IA
gains for OFDMA-based cellular systems.
\end{abstract}

\IEEEpeerreviewmaketitle

\section{Introduction}\label{section1}

This paper studies orthogonal frequency-division multiple access
(OFDMA) based cellular systems with universal frequency reuse, in
which adjacent cells share the same frequency band for simultaneous
transmission to improve the spectrum efficiency. However, for such
systems, the inter-cell interference (ICI) control becomes crucial,
which has recently drawn significant attention (see, e.g.,
\cite{Gesb10} and references therein). In general, there are two
approaches to cope with the ICI in multi-cell systems \cite{Gesb10}:
\emph{interference coordination} and \emph{network MIMO}
(multiple-input multiple-output). The former approach mitigates the
ICI via cooperative resource allocation across different cells based
on their shared channel state information (CSI), where the latter
approach utilizes the ICI via baseband-level signal cooperation for
joint encoding and/or decoding at the base stations (BSs). Although
promising from the viewpoint of theoretical performance, network
MIMO requires the baseband time synchronization as well as message
sharing among different BSs, which is challenging to implement for
existing cellular systems. As such, in this paper, we focus our
study on the interference coordination approach.

Existing resource allocation schemes (e.g.,
\cite{Vent09,Yu2010,Da2011ICC}) for multi-cell OFDMA systems have
adopted a \emph{subcarrier-separation} approach, whereby all
subcarriers (SCs) are treated as separate dimensions for user
selection and power allocation. However, a recent study
\cite{Cada2008Insepa} has revealed that for a \emph{parallel
interference channel} (PIC) consisting of parallel Gaussian
interference subchannels, joint precoding over parallel subchannels
improves the sum capacity over the case without precoding for some
specially designed channel realizations, i.e., the PIC is in general
\emph{non-separable}. Motivated by this finding, in this paper, we
study a new approach to design cooperative resource allocation for
multi-cell OFDMA, by exploiting frequency-domain precoding over
parallel SCs.

The precoding technique used for the PIC in \cite{Cada2008Insepa} is
known as \emph{interference alignment} (IA). By properly aligning
the interference at the receivers via linear precoding at the
transmitters, for $K$-user Gaussian interference channels, it is
shown in \cite{Cada2008Aug} that the IA technique achieves a
sum-rate multiplexing gain of $K/2$ per time, frequency or antenna
dimension. In other words, IA enables interference-free
communications for all the users provided that each user utilizes
only half of the available degrees of freedom (DoF). Although IA has
been largely investigated in cases with time-domain symbol extension
or with spatial beamforming via multiple antennas, how to exploit
frequency-domain IA to design efficient OFDMA resource allocation in
a multi-cell scenario remains open and has been rarely studied in
the literature, to our best knowledge.

In this paper, we investigate the problem of exploiting IA in
cooperative resource allocation in OFDMA-based cellular systems. For
the purpose of exposition, we study the OFDMA downlink transmission
in a simplified three-cell system with universal frequency reuse.
All the BSs and user terminals are assumed to be each equipped with
a single antenna, and the system is thus modeled by a \emph{SISO
(single-input single-output) interfering broadcast channel (BC)}.
With the objective of maximizing the system weighted sum-rate, the
joint optimization of frequency-domain precoding via IA, SC
scheduling with user selection, and power allocation is studied in
this paper. From the numerical experiments under a symmetric channel
setup (with unit average channel gain for all in-cell direct links
and the same average channel gain for all cross-cell interference
links), we find that the IA-based resource allocation scheme
demonstrates its advantages over the traditional scheme without
frequency-domain precoding only when the cross-cell link has a
comparable strength as the direct link, and the receiver
signal-to-noise (SNR) is sufficiently large. Motivated by this
observation, a hybrid scheme is proposed for practical cellular
systems with heterogenous channel conditions. In this hybrid scheme,
the total spectrum is divided into two subbands, over which the
IA-based scheme and the traditional scheme are applied for resource
allocation to users located in the \emph{cell-intersection region}
and \emph{cell-non-intersection region}, respectively. It is shown
by simulation results that this hybrid scheme can be flexibly
designed to exploit the downlink IA gains for OFDMA-based cellular
systems with cooperative interference control.

\begin{figure}[!t]
  \centering
  \includegraphics[width=80mm]{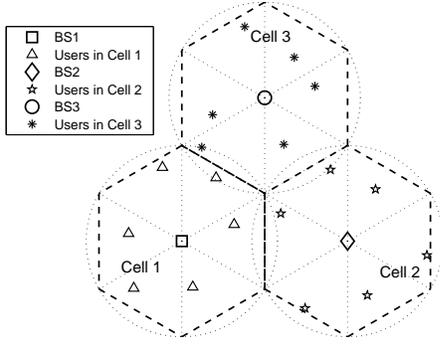} \\
\vspace{-12pt}
  \caption{System model for a cluster of three cells.}
  \label{fig.1}
\end{figure}

\section{System Model}\label{section2}

As shown in Fig. \ref{fig.1}, for the purpose of exposition, we
consider a three-cell system, in which all cells  share the same
frequency band for simultaneous downlink transmission. In each cell,
OFDMA is assumed for the downlink transmission, where the total
shared bandwidth is equally divided into $N$ SCs indexed by $n \in
\Lambda {\rm{ = }}\left\{ {1,2,...,N} \right\}$. The users in each
cell are scheduled for transmission over orthogonal SCs, while the
users' SC allocation are allowed to change from one scheduling
time-slot to another. A slow fading environment is assumed, in which
all the channels involved in the system are constant during each
scheduling slot, but can vary from slot to slot. In one particular
slot, each SC is assumed to be used by at most one user inside each
cell due to OFDMA, but allowed to be shared by users from different
cells due to universal spectrum sharing. We use $\Omega = \left\{
{1,2,3} \right\}$ to represent the set of three cells, and denote
the user associated with Cell $m$ as ${k_m} \in {\Delta _m}{\rm{ =
}}\left\{ {1,2,...,{K_m}} \right\}, m \in \Omega$, where $K_m$ is
the total number of users in Cell $m$. In addition, the complex
downlink baseband channel response  from the BS of Cell $j$ to user
$k_m$ in Cell $m$ at SC $n$ is denoted as $h_{j{k_m}}^n$. Each BS is
assumed to have a total transmit power constraint given by
$P_m^{{\rm{tot}}}, m \in \Omega$. Furthermore, we use the matrix
$\bf{\pi}$ of size $3\times N$ to denote the SC allocation in the
whole system, of which the $(m,n)$-th element, denoted by $\pi _m^n$
with $m\in\Omega, n\in\Lambda$, indicates the user assigned to SC
$n$ in Cell $m$.

Note that for all the resource allocation schemes presented in the
sequel, we assume that there exists a central controller that
collects all the CSI in the network via dedicated feedback channels
or backhaul networks, and is thus enabled to perform a centralized
resource allocation.

\section{Traditional Resource Allocation}\label{section3}

In this section, we present one traditional scheme for cooperative
resource allocation in the studied three-cell system, which is based
on the subcarrier-separation principle, i.e., all the SCs are
treated as separate dimensions for user selection and power
allocation. First, supposing that $\pi _m^n=k_m$, the baseband
signal received by user $k_m$ at SC $n$ is written as
\begin{equation}\label{equ.1}
y_{{k_m}}^n = h_{m{k_m}}^nx_m^n + \sum\limits_{j = 1,j \ne m}^3
{h_{j{k_m}}^nx_j^n}  + z_{{k_m}}^n,
\end{equation}
where $x_m^n$ is the complex symbol transmitted by BS $m$ at SC $n$,
${E}[{{{|{x_m^n}|}^2}}] = p_m^n$ with $p_m^n$ denoting the power
allocated, and $z_{{k_m}}^n$ is the circularly symmetric complex
Gaussian (CSCG) noise at the receiver with zero mean and variance
${\sigma ^2}$.

The signal-to-interference-plus-noise-ratio (SINR) of user $k_m$ in
Cell $m$ at SC $n$ is given by
\begin{equation}\label{equ.2}
SINR_{m{k_m}}^n = \frac{{p_m^ng_{m{k_m}}^n}}{{1 + \sum\limits_{j =
1,j \ne m}^3 {p_j^ng_{j{k_m}}^n} }},
\end{equation} where $g_{j{k_m}}^n = {{{{| {h_{j{k_m}}^n} |}^2}} \mathord{/
 {\vphantom {{{{| {h_{j{k_m}}^n}|}^2}} {{\sigma ^2}}}}
 \kern-\nulldelimiterspace} {{\sigma ^2}}}$. Then, the maximum achievable transmission rate at this SC is given by
(normalized to be in bps/Hz)
\begin{equation}\label{equ.3}
r_{m{k_m}}^n = \frac{1}{N}{\log _2}\left( {1 + SINR_{m{k_m}}^n}
\right),
\end{equation}
Hence, the problem of maximizing the weighted sum-rate of all users
in the system can be expressed as
\begin{equation}\label{equ.4}
\begin{array}{l}
 \mathop {\max }\limits_{{\bf{p}} \succeq 0,{\bf{\pi }} \in \Pi } {\rm{ }}U = \sum\limits_{n = 1}^N {\sum\limits_{m = 1}^3 {{\omega _{m{k_m}}}r_{m{k_m}}^n} }  \\
 s.t.{\rm{ }}\sum\limits_{n = 1}^N {p_m^n}  \le P_m^{{\rm{tot}}},{\rm{ }}m \in \Omega  \\
 \end{array}
\end{equation}
where ${\bf{p}} = \left\{ {{{\bf{p}}^1},...,{{\bf{p}}^N}} \right\}$
is a $3$-by-$N$ power allocation matrix consisting of column vectors
${{\bf{p}}^n} = {\left[ {p_1^n,p_2^n,p_3^n} \right]^T}, n \in
\Lambda$, $\Pi$ denotes the set of all possible SC allocation
matrices for ${\bf \pi}$, and ${\omega _{m{k_m}}}$ is the
(non-negative) weight of user $k_m$ in Cell $m$.

The problem in (\ref{equ.4}) can be shown to be non-convex over
${\bf{p}}$ due to the non-concave rate functions in (\ref{equ.3})
even for a fixed ${\bf \pi}$, and is thus in general non-convex with
arbitrary ${\bf \pi}$. Nevertheless, the Lagrange duality method can
be applied to this problem to obtain a set of close-to-optimal
solutions, which, as verified by our numerical experiments, usually
converge to the optimal solutions when the number of SCs becomes
large (i.e., the duality gap converges to zero as $N\rightarrow
\infty$). More specifically, the partial Lagrangian of problem
(\ref{equ.4}) is given by
\begin{equation}\label{equ.5}
L\left( {{\bf{p}},{\bf{\pi }},{\bf{\lambda }}} \right) = U +
{{\bf{\lambda }}^T}\left( {{{\bf{P}}^{\rm{tot}}} - \sum\limits_{n =
1}^N {{{\bf{p}}^n}} } \right),
\end{equation} where ${{\bf{P}}^{{\rm{tot}}}} = {\left[
{P_1^{{\rm{tot}}},P_2^{{\rm{tot}}},P_3^{{\rm{tot}}}} \right]^T}$ is
the power constraint vector, and ${\bf{\lambda }} = {\left[
{{\lambda _1},{\lambda _2},{\lambda _3}} \right]^T}$ is the vector
of non-negative dual variables. The Lagrange dual function then
becomes
\begin{equation}\label{equ.6}
f\left( {\bf{\lambda }} \right) = \mathop {\max }\limits_{{\bf{p}}
\succeq 0,{\bf{\pi }} \in \Pi } L\left( {{\bf{p}},{\bf{\pi
}},{\bf{\lambda }}} \right).
\end{equation} Thus, the dual problem can be defined as
\begin{equation}\label{equ.7}
\mathop {\min }\limits_{{\bf{\lambda }} \succeq 0} f\left(
{\bf{\lambda }} \right).
\end{equation}

For a given vector $\bf{\lambda}$, $f\left( {\bf{\lambda }} \right)$
can be simplified as
\begin{equation}\label{equ.8}
f\left( {\bf{\lambda }} \right) = \mathop {\max }\limits_{{\bf{p}}
\succeq 0,{\bf{\pi }} \in \Pi } \sum\limits_{n = 1}^N {{f_n}}  +
{{\bf{\lambda }}^T}{{\bf{P}}^{{\rm{tot}}}},
\end{equation} where
\begin{equation}\label{equ.9}
{f_n} = \sum\limits_{m = 1}^3 {\left( {{\omega _{{k_m}}}r_{m{k_m}}^n - {\lambda _m}p_m^n} \right)}.  
\end{equation}
The maximization problem in (\ref{equ.8}) is thus decoupled into
parallel per-SC based subproblems, which are given by
\begin{equation}\label{equ.10}
\mathop {\max }\limits_{{{\bf{p}}^n} \succeq 0,{{\bf{\pi }}^n} \in
{\Pi ^n}} {f_n},{\rm{ }}n \in \Lambda,
\end{equation} where ${{\bf{\pi }}^n} = {\left[ {\pi _1^n,\pi _2^n,\pi _3^n} \right]^T} \in {\Pi
^n}$ with ${\Pi ^n}$ denoting the $n$th column of ${\Pi}$. The
solutions to the problems in (\ref{equ.10}) have been studied in
e.g. \cite{Vent09,Yu2010,Da2011ICC} by various iterative methods;
the details are thus omitted for brevity. Last, subgradient-based
methods such as the ellipsoid method \cite{Da2011ICC} can be used to
iteratively search for the optimal ${\bf{\lambda }}$ in the dual
problem to make the corresponding power allocation solutions of the
problems in (\ref{equ.10}) satisfy all the per-BS power constraints.

Note that the above joint user SC and power allocation scheme is
based upon the premise that there is no frequency-domain precoding
over parallel SCs at each BS. Next, we will propose an IA-based
resource allocation scheme with frequency-domain precoding.

\section{IA-Based Resource Allocation}\label{section4}

In this section, we propose a new scheme for cooperative multi-cell
OFDMA downlink resource allocation with IA-based frequency-domain
precoding. Specifically, two non-adjacent SCs, denoted by
$(n_1,n_2)$, are grouped to form the $n$th SC-pair, which supports
simultaneous downlink transmission to three users, denoted by
$(k_1,k_2,k_3)$, each being selected from one of the three cells,
via linear transmit precoding over the two chosen SCs. Note that for
notational convenience, we use $n$ to denote the SC-pair index in
this section, where $n \in \tilde \Lambda {\rm{ = }}\left\{
{1,2,...,{N \mathord{\left/ {\vphantom {N 2}} \right.
\kern-\nulldelimiterspace} 2}} \right\}$ (assume that $N$ is an even
number).

Without loss of generality, we consider the baseband signals for
user $k_1$ over the $n$th SC-pair $(n_1,n_2)$, which are given by
\begin{equation}\label{equ.11.1}
y_{{k_1}}^{{n_1}} = h_{1{k_1}}^{{n_1}}x_1^{{n_1}} +
h_{2{k_1}}^{{n_1}}x_2^{{n_1}} + h_{3{k_1}}^{{n_1}}x_3^{{n_1}} +
z_{{k_1}}^{{n_1}},
\end{equation}
\begin{equation}\label{equ.11.2}
y_{{k_1}}^{{n_2}} = h_{1{k_1}}^{{n_2}}x_1^{{n_2}} +
h_{2{k_1}}^{{n_2}}x_2^{{n_2}} + h_{3{k_1}}^{{n_2}}x_3^{{n_2}} +
z_{{k_1}}^{{n_2}},
\end{equation}
where $x_m^{{n_1}} = v_m^{{n_1}}s_m^{{n}},x_m^{{n_2}} =
v_m^{{n_2}}s_m^{{n}},m \in \Omega $ with $v_m^{{n_1}},v_m^{{n_2}}$
being the precoder weights for user $k_m$ of Cell $m$ over the $n$th
SC-pair, and $s_m^n$ being the information-bearing symbol for user
$k_m$. Together, (\ref{equ.11.1}) and (\ref{equ.11.2}) can be
further expressed in the matrix form shown as follows:
\begin{equation}\label{equ.12}
{\bf{y}}_{{k_1}}^n = {\bf{H}}_{1{k_1}}^n{\bf{v}}_1^ns_1^n +
{\bf{H}}_{2{k_1}}^n{\bf{v}}_2^ns_2^n +
{\bf{H}}_{3{k_1}}^n{\bf{v}}_3^ns_3^n + {\bf{z}}_{{k_1}}^n,
\end{equation}
where ${\bf{v}}_m^n = {\left[ {v_m^{{n_1}},v_m^{{n_2}}} \right]^T}$
is the precoding vector for user $k_m$,
\begin{equation}\label{equ.12.H}
{\bf{H}}_{m{k_1}}^n = \left[ {\begin{array}{*{20}{c}}
   {h_{m{k_1}}^{{n_1}}} & 0  \\
   0 & {h_{m{k_1}}^{{n_2}}}  \\
\end{array}} \right],m \in \Omega,
\end{equation}
is the equivalent (diagonal) channel matrix from BS $m$ to user
$k_1$, and ${\bf{z}}_{{k_1}}^n$ is the noise vector for user $k_1$,
all defined for the $n$th SC-pair.  Based on the analysis for user
$k_1$, the downlink transmission from the three BSs to their
respective users over the $n$-th SC pair can be represented by
\begin{equation}\label{equ.13}
{\bf{y}}_{{k_m}}^n = {\bf{H}}_{m{k_m}}^n{\bf{v}}_m^ns_m^n +
\sum\limits_{j = 1,j \ne m}^3 {{\bf{H}}_{j{k_m}}^n{\bf{v}}_j^ns_j^n}
+ {\bf{z}}_{{k_m}}^n,
\end{equation}
$ m\in\Omega, n\in\tilde{\Lambda}$. At the receiver of user $k_m$,
an interference suppression vector ${\bf{u}}_m^n = \left[
{u_m^{{n_1}},u_m^{{n_2}}} \right]$ is pre-multiplied with the
received signal to eliminate the inter-cell interference, which
leads to the following equivalent channel model:
\begin{equation}\label{equ.14}
\tilde y_{{k_m}}^n = {\bf{u}}_m^n{\bf{y}}_{{k_m}}^n = \tilde
h_{m{k_m}}^ns_m^n + \sum\limits_{j = 1,j \ne m}^3 {\tilde
h_{j{k_m}}^ns_j^n}  + \tilde z_{{k_m}}^n,
\end{equation}
where
\begin{equation}\label{equ.14.h}
\tilde h_{m{k_m}}^n = {\bf{u}}_m^n{\bf{H}}_{m{k_m}}^n{\bf{v}}_m^n,
\end{equation}
\begin{equation}\label{equ.14.RI}
\tilde h_{j{k_m}}^n = {\bf{u}}_m^n{\bf{H}}_{j{k_m}}^n{\bf{v}}_j^n, j
\in \Omega, j \ne m,
\end{equation}
\begin{equation}\label{equ.14.z}
\tilde z_{{k_m}}^n = {\bf{u}}_m^n{\bf{z}}_{{k_m}}^n.
\end{equation}
Note that (\ref{equ.14}) bears a similar expression as
(\ref{equ.1}), with $\tilde h_{m{k_m}}^n, \tilde h_{j{k_m}}^n$ being
equivalent channel gains for direct and inter-cell links,
respectively. Assuming that ${\bf{v}}_m^n,{\bf{u}}_m^n, m \in
\Omega, n\in\tilde{\Lambda}$ are all of unit norm and
${E}[|s_m^{{n}}|^2]=\tilde{p}_m^n$, the achievable rate over the
$n$th SC-pair for user $k_m$ is given by
\begin{equation}\label{equ.15}
\tilde r_{m{k_m}}^n = \frac{1}{N}{\log _2}\left( {1 + \frac{{{\tilde
p}_m^n\tilde g_{m{k_m}}^n}}{{1 + \sum\limits_{j = 1,j \ne m}^3
{\tilde{p}_j^n\tilde g_{j{k_m}}^n} }}} \right),
\end{equation} where $\tilde g_{m{k_m}}^n = {{{{| {\tilde h_{m{k_m}}^n}|}^2}}
\mathord{{\vphantom {{{{| {\tilde h_{m{k_m}}^n} |}^2}} {{\sigma
^2}}}} /\kern-\nulldelimiterspace} {{\sigma ^2}}}$.

Since searching for the optimal SC-pairs over $\Lambda$ is
computationally prohibitive, we use a pre-determined SC-pairing
method as follows: for the $n$th SC-pair, the two SCs are given by
$n_1=n, n_2=n+N/2$. This paring method is to maximize the frequency
gap between the two SCs in each pair, so as to maximally exploit the
frequency-domain channel diversity. With the above SC-pairs, the
weighted sum-throughput of the three-cell system is obtained by
solving the following problem:
\begin{equation}\label{equ.16}
\begin{array}{l}
 \mathop {\max }\limits_{{\bf{\tilde p}} \succeq 0,{\bf{\tilde \pi }} \in \tilde \Pi } {\rm{ }}\tilde U = \sum\limits_{n = 1}^{{N \mathord{\left/
 {\vphantom {N 2}} \right.
 \kern-\nulldelimiterspace} 2}} {\sum\limits_{m = 1}^3 {{{\omega }_{m{k_m}}}\tilde r_{m{k_m}}^n} }  \\
 s.t. {\ } \sum\limits_{n = 1}^{{N \mathord{\left/
 {\vphantom {N 2}} \right.
 \kern-\nulldelimiterspace} 2}} {{\tilde p}_m^n}  \le P_m^{{\rm{tot}}},{\rm{ }}m \in \Omega  \\
 \end{array}
\end{equation}
where ${\bf{\tilde p}} = \left\{ {{{\bf{\tilde
p}}^1},...,{{\bf{\tilde p}}^{{N \mathord{\left/ {\vphantom {N 2}}
\right. \kern-\nulldelimiterspace} 2}}}} \right\}$ with ${\bf{\tilde
p}}^n=[\tilde{p}_1^n,...,\tilde{p}_3^n]^T, n\in\tilde{\Lambda}$, and
${\bf{\tilde \pi}}$ is a $3$-by-$N/2$ matrix with its $n$th column
vector ${{\bf{\tilde \pi }}^n} \in {\tilde \Pi ^n}$ specifying the
users $(k_1,k_2,k_3)$ selected from the three cells for the $n$th
SC-pair. Note that ${\tilde \Pi}$ and ${\tilde \Pi^n}$ are similarly
defined as $\Pi$ and $\Pi^n$ in Section III, respectively, while it
is worth noting that $n$ here refers to the SC-pair index instead of
the SC index in Section III.

Similarly as the derivation given in Section III, the problem in
(\ref{equ.16}) can be decoupled into parallel per-SC-pair based
resource allocation subproblems in the dual domain, i.e.,
\begin{equation}\label{equ.18}
\mathop {\max }\limits_{{{\bf{\tilde p}}^n} \succeq 0,{{{\bf{\tilde
\pi }}}^n} \in {{\tilde \Pi }^n}} {\tilde f_n} = \sum\limits_{m =
1}^3 {\left( {{{\omega }_{m{k_m}}}\tilde r_{m{k_m}}^n - {{\tilde
\lambda }_m}\tilde{p}_m^n} \right)} ,{\rm{ }}n \in \tilde \Lambda,
\end{equation} where ${\bf{\tilde \lambda }} = {\left[ {{{\tilde \lambda }_1},{{\tilde \lambda }_2},{{\tilde \lambda }_3}}
\right]^T}$ is the vector of non-negative dual variables associated
with the per-BS power constraints.

Consider first the case when perfect IA is achievable, i.e., all the
inter-cell interference gains in (\ref{equ.14.RI}) become zero. In
this case, the residual interference (RI) terms in the denominator
of (\ref{equ.15}) vanish, which results in decoupled user power
optimization in (\ref{equ.18}) for each SC-pair. Specifically, by
letting
\begin{equation}\label{equ.19}
\frac{{\partial {{\tilde f}_n}}}{{\partial \tilde{p}_m^n}} =
\frac{{{{\omega }_{m{k_m}}}}}{N}\frac{{\tilde g_{m{k_m}}^n}}{{\ln
2\left( {1 + \tilde{p}_m^n\tilde g_{m{k_m}}^n} \right)}} - {\tilde
\lambda _m} = 0,
\end{equation}
we obtain the following optimal power allocation:
\begin{equation}\label{equ.20}
\tilde{p}_m^n = {\left( {\frac{{{{\omega }_{m{k_m}}}}}{{{{\tilde
\lambda }_m}N\ln 2}} - \frac{1}{{\tilde g_{m{k_m}}^n}}} \right)^ +
}, m\in\Omega, n\in\tilde{\Lambda},
\end{equation}
which resembles the conventional ``water-filling'' solution.

However, to our best knowledge, no closed-form expressions for
${\bf{v}}_m^n$ and ${\bf{u}}_m^n$ that achieve the perfect IA have
been found for our studied scenario. Thus, we resort to the existing
``Distributed Interference Alignment'' algorithm in \cite{Goma2008}
to compute the transmitter precoding and receiver interference
suppression vectors given the CSI for any set of three users and
SC-pair. Based on our numerical experiments, it is found that with
this algorithm, small but non-zero RI usually exists for the
equivalent three-user $2\times 2$ MIMO interference channel given by
(\ref{equ.13}), probably due to the particular structure of diagonal
direct-link and cross-link channel matrices. In this case with the
RI, the problem in (\ref{equ.16}) can be solved similarly as that in
(\ref{equ.4}) in the case of traditional scheme with virtually $N/2$
instead of $N$ SCs. For brevity, the details are omitted.

\begin{figure}[!t]
  \centering
  \includegraphics[width=80mm]{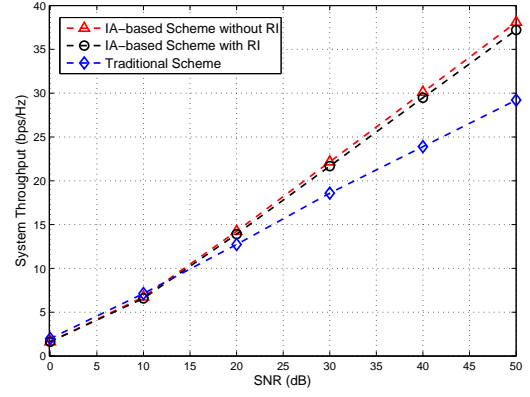} \\ \vspace{-6pt}
  \caption{System throughput under a symmetric channel setup with $h= 1$.}
  \label{fig.2}  \vspace{-6pt}
\end{figure}

\begin{figure}[!t]
  \centering
  \includegraphics[width=80mm]{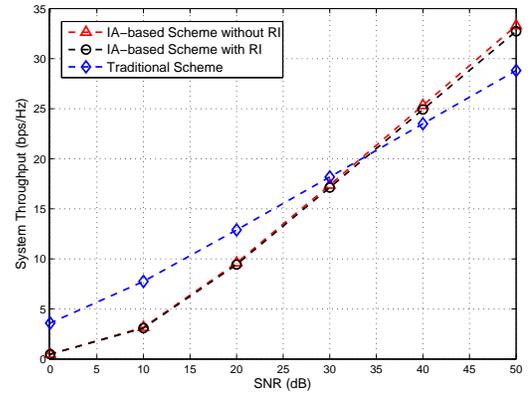} \\ \vspace{-6pt}
  \caption{System throughput under a symmetric channel setup with $h=0.1$.}
  \label{fig.3} \vspace{-8pt}
\end{figure}

Next, we compare the performance of the tradition resource
allocation scheme in Section III and the newly proposed IA-based
scheme in terms of the system sum-rate (i.e., ${\omega_{m{k_m}}}=1,
\forall m\in\Omega, k_m\in\Delta_m$). In order to gain intuition
from the numerical results, we assume an ideal symmetric channel
model for the three-cell system in Fig. \ref{fig.1}, in which
$h_{j{k_m}}^n$'s are modeled as independent CSCG random variables
$\sim \mathcal{CN}\left( {0,1} \right)$ for $j=m$, and $\sim
\mathcal{CN}\left( {0,h} \right)$ for $j\neq m$, respectively. In
addition, it is assumed that $N=64$, $K_m=4, m\in\Omega$,
$P_m^{{\rm{tot}}}$'s are same for all three BSs, and $\sigma^2=1$.

Fig. \ref{fig.2} shows the sum-rate versus the average direct-link
SNR at each user's receiver with $h=1$, i.e., the direct-link has
the same average power gain as the cross-link in the symmetric
channel setup. It is observed that the IA-based scheme with the RI
performs slightly worse than that without the RI (assuming perfect
IA and thus neglecting the RI terms in (\ref{equ.15})). In addition,
the IA-based scheme is observed to perform better than the
traditional scheme when SNR is larger than 15dB, and eventually
achieve a DoF gain of $3/2$ compared to the traditional scheme when
SNR goes to infinity. This DoF gain can be explained as follows: For
the IA-based scheme (without RI), three data streams are transmitted
without mutual interference over each SC-pair consisting of two SCs,
while for the traditional scheme, when the SNR is sufficiently high,
the optimal SC allocation tends to be orthogonal, i.e., each SC is
assigned to only one user from a particular cell, while no users
from the other two cells are assigned to the same SC.

Similar observations are obtained in Fig. \ref{fig.3} for the
symmetric channel setup with $h=0.1$. However, in this case with
weak cross-link interference, the SNR threshold beyond which the
IA-based scheme outperforms the traditional scheme increases from
15dB in the previous case of $h=1$ to 35dB in the present case. In
addition, the IA-based scheme is observed to perform notably worse
than the traditional scheme from low to moderate SNRs (0-20dB) that
are typical for cellular systems. This performance gap is because
the IA-based scheme always transmits three data streams over two SCs
regardless of the SNR, while the traditional scheme can support up
to 3 users each from one cell at any single SC when the inter-cell
interference is sufficiently weak as in the present case.

\section{Hybrid Scheme}\label{section5}

\begin{figure}[!t]
  \centering
  \includegraphics[width=78mm]{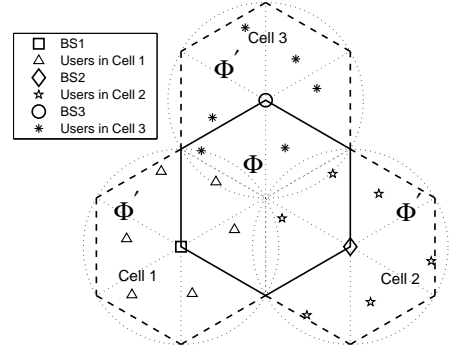} \\
  \caption{Cell and subcarrier partition for the hybrid scheme.}
  \label{fig.4}\vspace{-6pt}
\end{figure}

The numerical results for the symmetric channel setup reveal that
the IA-based scheme with frequency-domain precoding provide
throughput gains over the traditional scheme without precoding only
when the cross-link has a comparable strength with the direct-link
and the operating SNR is sufficiently high. The above conditions are
rarely satisfied in practical cellular systems due to randomized
user locations and distance-dependent signal attenuation. This
motivates our following hybrid scheme for resource allocation in a
three-cell system with heterogenous channel conditions (cf. Fig.
\ref{fig.4}): For the users located within the six sectors (two from
each cell) that lie in the intersection region of the three cells,
namely \emph{cell-intersection region} (CIR), a dedicated set of
SCs, denoted by $\Phi$, is reserved over which the IA-based scheme
is applied, since in this region the aforementioned operating
conditions for IA are more likely to be satisfied; in contrast, for
the rest of the users located in the \emph{cell-non-intersection
region} (CNIR), the remaining set of SCs, denoted by $\Phi'$, is
allocated over which the traditional scheme is applied. For example,
we can design the sizes of $\Phi$ and $\Phi'$ to be proportional to
the number of sectors in the CIR and CNIR, respectively, i.e., $\Phi
= \left\{ {1,2,...,\left\lfloor {{N \mathord{\left/
 {\vphantom {N 6}} \right.
 \kern-\nulldelimiterspace} 6}} \right\rfloor ,...,1 + \left\lfloor {{N \mathord{\left/
 {\vphantom {N 2}} \right.
 \kern-\nulldelimiterspace} 2}} \right\rfloor ,} \right.$ $\left. {2 + \left\lfloor {{N \mathord{\left/
 {\vphantom {N 2}} \right.
 \kern-\nulldelimiterspace} 2}} \right\rfloor ,...,\left\lfloor {{{2N} \mathord{\left/
 {\vphantom {{2N} 3}} \right.
 \kern-\nulldelimiterspace} 3}} \right\rfloor } \right\}$, and thus $\Phi^\prime = \Lambda  - \Phi$.

Similar to the problems formulated in (\ref{equ.4}) and
(\ref{equ.16}), the following problem maximizes the system weighted
sum-rate in the case of hybrid scheme (thus similarly solvable):
\begin{eqnarray*}
\mathop {\max }\limits_{{{\bf{p}}_h} \succeq 0,{{\bf{\pi }}_h} \in
{\Pi _h}} {U_h} = \sum\limits_{n \in \tilde \Phi }^{}
{\sum\limits_{m = 1}^3 {{\omega _{m{k_m}}}\tilde r_{m{k_m}}^n} }  +
\sum\limits_{n \in \Phi '}^{} {\sum\limits_{m = 1}^3 {{\omega
_{m{k_m}}}r_{m{k_m}}^n} }
\end{eqnarray*}
\begin{equation}\label{equ.23}
s.t.{\rm{ }}\sum\limits_{n \in \tilde \Phi }^{} {\tilde{p}_m^n}  +
\sum\limits_{n \in \Phi '}^{} {p_m^n}  \le P_m^{{\rm{tot}}},{\rm{
}}m \in \Omega
\end{equation} where ${r_{m{k_m}}^n}$ and ${\tilde r_{m{k_m}}^n}$
are given in (\ref{equ.3}) and (\ref{equ.15}) respectively; the
power and SC allocations are similarly defined as
${{\bf{p}}_h},{{\bf{\pi }}_h}$, and $\tilde \Phi  = \left\{
{1,2,...,\left\lfloor {{N \mathord{\left/ {\vphantom {N 6}} \right.
\kern-\nulldelimiterspace} 6}} \right\rfloor } \right\}$ is used to
index the SC-pairs in $\Phi$.

\begin{figure}[!t]
  \centering
  \includegraphics[width=80mm]{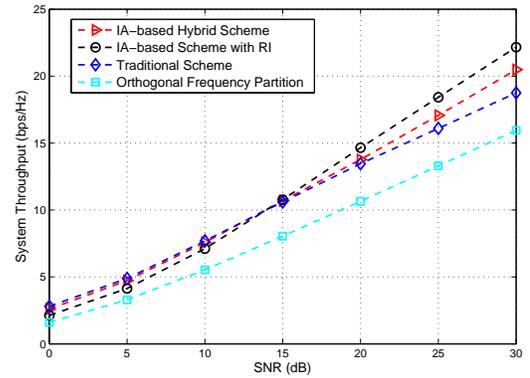} \\
  \caption{System throughput of a three-cell system with heterogenous channel conditions.}
  \label{fig.5}\vspace{-6pt}
\end{figure}

Fig. \ref{fig.5} shows the system sum-rates for various schemes in a
three-cell system with heterogenous channels with distance-dependent
attenuation (the attenuation factor is two). One additional
traditional scheme, namely \emph{orthogonal frequency partition}
(OFP), is also considered, which allocates each cell $1/3$ of the
total spectrum for orthogonal downlink transmission. It is assumed
that $N=256$, $K_m=12, m\in\Omega$, and the SNR shown in the figure
is the average SNR measured at the cell edge. It is observed that
the hybrid scheme can exploit the advantages of both traditional and
IA-based resource allocation schemes for low and high SNRs,
respectively, and thus achieve an overall more balanced performance.
It is also worth noting that by changing the relative sizes of
$\Phi$ and $\Phi'$, alternative designs of the hybrid scheme can be
flexibly obtained to maximize the throughput for a given SNR value.

\section{Conclusion}

This paper studies the downlink cooperative interference control in
cellular OFDMA systems. A new IA-based resource allocation scheme is
proposed, which jointly optimizes the frequency-domain precoding,
subcarrier user selection, and power allocation to maximize the
system throughput. For practical cellular systems with heterogenous
channel conditions, a hybrid scheme is proposed to exploit the
downlink IA gains.

\end{document}